# Mapping Student-AI Interaction Dynamics in Multi-Agent Learning Environments: Supporting Personalised Learning and Reducing Performance Gaps

Zhanxin Hao[1]; Jie Cao[2]; Ruimiao Li[3]; Jifan Yu[1]; Zhiyuan Liu[4]; Yu Zhang[1]*

1 Institute of Education, Tsinghua University, Beijing, China
2 School of Education, University of Pittsburgh
3 Institute of Education, Shanghai Normal University
4 Department of Computer Science and Technology, Tsinghua University, Beijing, China

**Abstract**

Multi-agent AI systems, which simulate diverse instructional roles such as teachers and peers, offer new possibilities for personalized and interactive learning. Yet, student-AI interaction patterns and their pedagogical implications remain unclear. This study explores how university students engaged with multiple AI agents, and how these interactions influenced cognitive outcomes (learning gains) and non-cognitive factors (motivation, technology acceptance). Based on MAIC, an online learning platform with multi-agent, the research involved 305 university students and 19,365 lines of dialogue data. Pre- and post-test scores, self-reported motivation and technology acceptance were also collected. The study identified two engagement patterns: co-construction of knowledge and co-regulation. Lag sequential analysis revealed that students with lower prior knowledge relied more on co-construction of knowledge sequences, showing higher learning gains and post-course motivation. In contrast, students with higher prior knowledge engaged more in co-regulation behaviors but exhibited limited learning improvement. Technology acceptance increased across all groups. These findings suggest that multi-agent AI systems can adapt to students' varying needs, support differentiated engagement, and reduce performance gaps. Implications for personalized system design and future research directions are discussed.

## 1. Introduction

Online learning has increasingly become a crucial mode of education. The advent of large language models (LLMs) has significantly advanced interactive, real-time conversation-based learning, enabling students to learn with a high degree of autonomy and address interactive deficiencies and

---

[1] Corresponding details: zhangyu2011@tsinghua.edu.cn

This work was funded by the Beijing Educational Science Foundation of the Fourteenth 5-year Planning (BAEA24024).

isolated learning challenges that have long hindered traditional online education (Kim, Chay, et al., 2024; Xu et al., 2024). Compared to traditional online learning environments (e.g., MOOCs), LLM-based online learning systems can support highly personalized learning experiences tailored to individual needs (Chen et al., 2024; Razafinirina et al., 2024). Meanwhile, the innovation demonstrates considerable promise in facilitating co-regulated learning processes between students and AI agents, which contributes to more effective attainment of learning goals.

In recent educational practice, a growing number of LLM-based agents have become key learning partners in online learning. They can function as personalized tutors for problem-solving (Albdrani & Al-Shargabi, 2023), teachable agents for programming education and debugging (Liu et al., 2024), generative agents for teacher training (Lu et al., 2024), and assessors for student evaluation (Willert & Würz, 2025). These applications demonstrate that intelligent agents perform well in playing roles, understanding human instructions, facilitating interaction, and making decisions, giving guidance which might enhance learning efficiency and ultimately improve their learning outcomes.

Building on the promising capabilities of individual LLM-based agents, the concept of LLM-based multi-agents has been proposed to leverage the collective intelligence and specialized skills of multiple agents. Compared to single-agent systems, multi-agent systems have the following advantages: 1) they can simulate the entire teaching process; 2) they provide a sense of presence through multi-role interactions; and 3) they support personalized learning as students could tailor the agent according to their needs (Zhang et al., 2024). Several studies have explored possible applications of AI-simulated environments where multiple agents work collaboratively and each assigns specific identities. For example, a research team from Stanford University has created a simulated town Smallville, where multiple agents play different roles to simulate real social activities (Park et al., 2023). In the field of education, scholars have begun to explore the potential applications of multi-agent systems. For example, researchers have explored a simulated classroom where different AI agents play the roles of both teachers and students, leading to more immersive and dynamic learning experiences (Yu et al., 2024). In another work, researchers constructed a multi-agent debating system, where agents with diverse perspectives or areas of expertise engage in debates on specific topics, exchanging viewpoints to enrich understanding (Hu et al., 2024). These developments indicate that LLM-based multi-agent systems could foster more interactive and personalized learning environments.

However, existing educational research mostly approaches the design of AI-supported learning systems from a technological perspective or focuses on evaluating the effectiveness of the usage of specific AI tools. A significant gap remains in understanding the nuanced process of student-AI interaction, particularly within a multi-agent learning environment. To address this gap, the present study systematically analyzes the dialogic interaction between students and different AI agents. Each interaction is meticulously coded to identify its nature, followed by transition analysis and lag sequential analysis (LSA) to uncover the underlying interaction dynamics. This research aims to explore novel patterns of learning engagement that arise from student-AI interactions and provide insights into the potential applicability of multi-agent learning systems in authentic educational settings. Specifically, this study addresses the following research questions:

1) What distinctive behavioral patterns of learning engagement emerge when students interact with multiple AI agents?

2) How do interaction patterns differ between students with high prior knowledge and those with low prior knowledge?

3) How does the interaction with AI agents contribute to students' learning gains?

**2. Literature Review**

**2.1 Interaction in online learning**

Online interaction has consistently been a key focus in the field of online learning, as it is closely linked to learner engagement and the overall quality of learning outcomes. Previous research has indicated that students who engage in the interaction are more likely to conduct effective educational transactions and thus may benefit from online learning (Gokool-Ramdoo, 2008; Yu et al., 2020).

Researchers have employed various methods to investigate interaction patterns in online learning environments. Traditional approaches have primarily utilized questionnaires (Kuo et al., 2014; Wang et al., 2022) and interviews (Arbaugh & Benbunan-Fich, 2007; Donnelly, 2010; Marks, Sibley, & Arbaugh, 2005; Rhode, 2009). Some studies have analyzed log data to identify specific behavioural patterns, leveraging statistical and machine learning techniques to examine their association with students' academic performance (Agudo-Peregrina et al., 2014). In addition, some studies have coded online behavioral or textual data and subsequently applied LSA to examine transitions between behaviors or states within online learning environments (Huang et al., 2019; Yang et al., 2018).

Moore (1989) divided the interaction in online learning into three types: student-student interaction, student-instructor interaction, and student-content interaction. Student-student interaction refers to the two-way communication among learners, such as exchanging ideas with classmates, discussing and receiving feedback from each other. Student-instructor interaction refers to the two-way communication between students and instructors, such as asking questions and getting feedback. Student-content interaction describes the engagement with course materials that help students actively make sense of the subject matter. Building on Moore's framework, Hillman, Willis, and Gunawardena (1994) recognized a need to introduce the fourth type of interaction: student-interface interaction. They argued that the advanced technological mediation of online learning environments necessitates an understanding of how students interact with the tools and platforms used for learning. They emphasized that students' success is influenced by their proficiency with the interface, including navigating learning tools and locating relevant information. In a later work, Soo and Bonk (1998) expanded Moore's framework by adding self-interaction, which refers to the self-regulation and reflexive thinking processes that are essential in self-directed e-learning. Unlike other interaction types, self-interaction does not naturally generate data in virtual learning environments, making it less frequently studied.

In terms of the impact of interaction, numerous studies have investigated how different interaction types affect various aspects of online learning, yielding mixed results. Firstly, student-instructor interaction is highly valued and is often associated with positive perceptions among learners, though in traditional online learning contexts, student-instructor interaction is costly and the least scalable as it requires the instructors' presence (Anderson, 2003; Moore, 1989). Arbaugh and Benbunan-Fich (2007) identified that student-instructor and student-system interactions are major predictors of perceived learning in their hierarchical modeling of online courses. In a later study, Kuo et al. (2014) demonstrated that student-instructor interaction significantly predicts students' learning satisfaction, reinforcing the notion that direct engagement with instructors enhances student experiences and outcomes. Student-interface interaction, which involves students' operational activities on learning platforms, is typically recorded in log data, such as login and logout times, course enrollments, calendar edits, forum participation, and contact additions (Joksimović et al., 2015). In adaptive learning systems, the dimension of interaction extends beyond simple operations; it includes feedback mechanisms through dashboards (Molenaar, 2022) and help-seeking behaviors, such as using hints (Goldin et al., 2013). Research by Joksimović et al. (2015) found that the time spent on student-interface interactions had a consistent and positive effect on learning outcomes, suggesting that the more students engage with the system, the better they perform academically.

While the student-student interaction within online learning environments has also been explored extensively, findings regarding its effectiveness remain mixed. On the one hand, peer interaction has been associated with improved learning outcomes as well as greater and diverse engagement (Kurucay & Inan, 2017; Lai et al., 2019; Huang, et al., 2019). For example, Yang et al. (2018)'s research indicated that the interaction patterns vary significantly among groups with different levels of engagement in online cooperative learning contexts. Groups with higher engagement levels displayed more frequent and sustained behaviors of negotiation and knowledge co-construction. On the other hand, research also suggests that students do not always perceive such interactions as beneficial, due to challenges like uneven participation, limited feedback, and the need for high self-regulation (Kuo et al., 2014; Thurmond et al., 2002; Rhode, 2009).

**2.2 Extending Online Interaction Paradigms: The Role of AI in Online Learning**

The emergence of generative AI, as represented by ChatGPT, has fundamentally transformed the way students interact with AI systems. Unlike earlier rule-based or menu-driven interfaces, generative AI supports natural, flexible, and text-based conversations, lowering the technical threshold for meaningful engagement. Recent research has identified a range of educational uses of generative AI, such as brainstorming, exploring new ideas, and problem-solving (Baidoo-Anu & Ansah, 2023; Pesovski et al., 2024; Šedlbauer, et al., 2024), thereby broadening the scope of AI applications and introducing new paradigms of student-AI interaction.

Investigations into the interactions between students and generative AI have begun to shed light on the distinct characteristics and outcomes of integrating AI in educational settings. A growing body of literature has focused on the functionalities of generative AI, such as ChatGPT, in assisting student learning. Recent reviews have categorized student-AI interactions into four predominant themes based on the roles and functions of AI: (1) authoritative resource, providing accurate and comprehensive responses; (2) facilitator, supporting self-directed learning by guiding learners through reflective processes; (3) collaborator, fostering collaborative dialogues that promote knowledge co-creation; and

(4) extended cognitive capacity, enhancing students' intellectual engagement. These roles have been linked to positive outcomes across various domains, including language learning (Tai & Chen, 2024; Chen & Pan, 2022), programming (Denny et al., 2022; Yilmaz & Yilmaz, 2023), and affective-motivational dimensions such as interest and engagement (Du & Daniel, 2024; Liang & Hwang, 2023).

However, much of this work focuses on the functional roles of a single generative AI agent in assisting learners, often in task-specific or short-term settings. Fewer studies delve into the interactional nuances—how students engage with AI over time, how they shift roles in dialogic processes, and how they coordinate meaning-making across different learning contexts. Urban et al. (2024), for instance, found that students perceive ChatGPT interactions as natural and low-pressure, resembling communication with a human teacher but without the associated social risks. Chen et al. (2025) further noted task-dependent differences in student-AI interaction, revealing that help-seeking with AI tends to be more pragmatic and non-linear, while interactions with human experts follow a more linear and theoretically guided trajectory.

To summarize, while current literature lays a strong foundation for understanding learner-AI interaction, research has largely centered on dyadic interactions between a student and a single AI agent. In contrast, multi-agent AI systems—in which a student interacts with multiple AI agents playing diverse pedagogical roles (e.g., peer, tutor, facilitator)—remain underexplored. This presents a critical opportunity to extend our understanding of online interaction paradigms. By examining how students navigate multi-agent dialogues, respond to differentiated agent roles, and construct meaning across AI-mediated exchanges, this study seeks to illuminate new dimensions of student-AI interaction that go beyond prior single-agent frameworks. Ultimately, this work aims to contribute to the development of more pedagogically grounded, interaction-rich AI-supported learning environments.

**3. Methodology**

**3.1 Settings: Massive AI-empowered Course System (MAIC)**

This study was conducted using a massive AI-empowered course system (MAIC). The MAIC system is an online learning platform that constructs a series of LLM-driven agents to support both teaching and learning (Yu et al., 2024). Multiple agents work collaboratively to facilitate the entire process of course design, instruction, and management. The interface agents in MAIC system include teacher and student agents, while the back-end agents consist of analytical agents such as the Classroom Situation Analysis Agent, Learning Situation Analysis Agent, and Classroom Director Agent. For course design the instructors upload relevant course materials, and the agents utilize a range of models such as multimodal understanding and knowledge structure extraction to draft teaching slides and lecture scripts. The instructor could further collaborate with agents to optimize the slides and scripts. Once the preparation is completed, the course is conducted by AI teaching agents which utilize highly human-like natural speech, closely mimicking the tone and delivery of real classroom teacher. Students listened the lecture while interacting with multiple AI peers. Four types of AI peers are designed to provide students with a more simulated learning environment-- Sparker, Questioner, Thinker and Notetaker--each with distinct functions as shown in Figure 1. Various analytical agents conduct analyses of the classroom proceedings and learning conditions to make decisions regarding the pacing of the class and which specific agent should appear to interact with students at a given moment.

**Figure 1**

*The classroom setup interface*

To study via MAIC, students begin by selecting the course they wish to study, which directs them to the classroom setup interface, as shown in Figure 1. In this interface, students can customize roles of classroom participants and select a learning mode, opting for either observation mode or interaction mode. In the observation mode, students watch the presentation of slides and listen to the AI teacher's explanation, meanwhile watching the interaction between AI peers and AI teachers. In contrast, interaction mode allows students to send messages at any time and receive responses from various AI agents. Students can interchange the two modes anytime they want. Additionally, students can adjust the intensity of classroom discussion, the intervals between speaking turns, and the speed of text display to tailor the learning experience to their preferences.

Once these settings are completed, learners enter the classroom, as shown in Figure 2. The left side of the interface presents the lecture slides, while the right is the interaction panel where students can chat with multiple agents.

**Figure 2**

*The classroom interface*

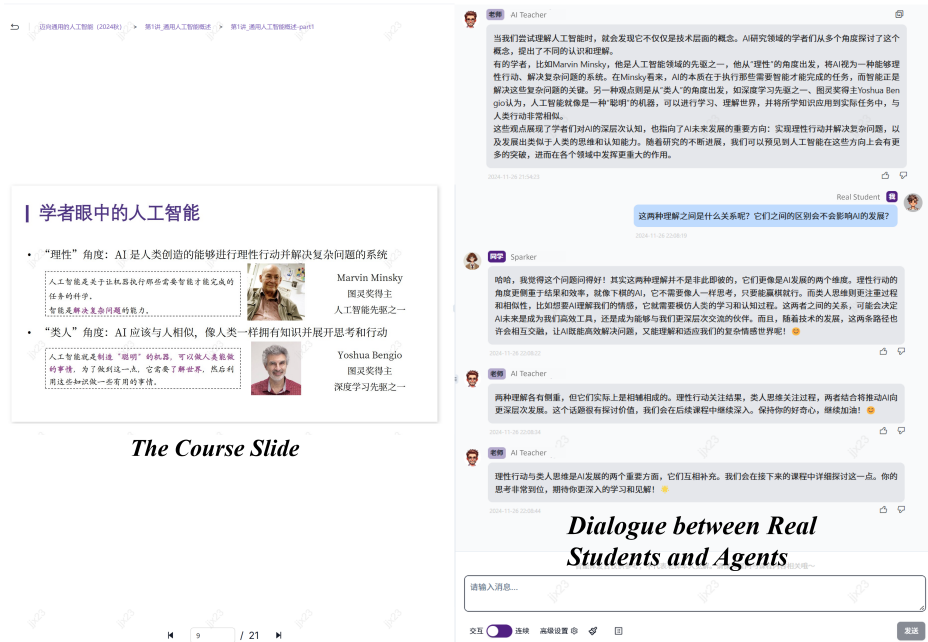

*The Course Slide*

*Dialogue between Real Students and Agents*

This study focuses on one introductory course entitled *Towards Artificial General Intelligence* on the MAIC platform. The course consists of six modules, including: 1) Overview of General Artificial Intelligence; 2) Fundamentals of Neural Networks and Large Models; 3) Large Models Integrating Visual, Language, and Sound Inputs; 4) Autonomous Agents; 5) AI + X; 6) AI Safety and Ethics. This course was also provided on one of the MOOC platforms.

**3.2 Participants**

A total of 305 students (34.75% female, $M_{age}$= 20.16, SD=1.26) from an elite university in China initially enrolled in the course via the MAIC platform. Additionally, forty-seven students who accessed the course through the MOOC platform, were also recruited to participate. Participants came from various academic disciplines and volunteered to take part in the study. Ethical approval was obtained from [UNIVERSITY NAME, anonymized for peer review), and consent forms were signed by all participants prior to the study. Students were asked to complete the course within two months. By the end of the course, a total of 131 students (30.53% female, $M_{age}$= 20.06, SD=1.26) had completed the entire course on the MAIC platform and 46 students completed the course on the MOOC platform.

**3.3 Data collection**

This study collected and analyzed two types of data: questionnaire data collected before and after the course, pre-test and post-test data, as well as dialogue data between students and AI agents during the course.

**3.3.1 Questionnaires**

The questionnaires were employed to gather students' learning motivation and attitude towards AI, which are possible associated with students' learning engagement. All self-reported measures were designed using well-established scales with demonstrated reliability.

**Learning Motivation:** Self-determined motivation is broadly defined as the degree to which individuals engage in an activity out of personal choice and intrinsic enjoyment (Blais, Sabourin, Boucher, & Vallerand, 1990; Grolnick & Ryan, 1987; Vallerand & Bissonnette, 1992). In this study, students' motivation for academic studies was assessed using a 12-item scale adapted from the Academic Motivation Scale (AMS) employed in prior research (e.g., Guay & Vallerand, 1997; Knapp et al., 2010; Sheldon, Osin, Gordeeva, Suchkov, & Sychev, 2017; Vansteenkiste, Lens, De Witte, & Deci, 2004; Weinstein & Ryan, 2010). The scale comprises two subscales: self-determined motivation and non-self-determined motivation, with six items assigned to each subscale. Responses were used to measure the extent of intrinsic versus extrinsic motivational factors influencing academic work. Cronbach's α for this scale is 0.82.

**AI technology acceptance:** Students' attitudes toward artificial intelligence (AI) stem from the broader concept of technology acceptance, which can be defined as "an individual's psychological state regarding their voluntary and intentional use of a particular technology" (Hendrick et al., 1984). This study employs the UTAUT2 (Unified Theory of Acceptance and Use of Technology) model to examine such attitudes. 30 items in the scale encompass eight core dimensions: Performance Expectancy, Effort Expectancy, Social Influence, Facilitating Conditions, Hedonic Motivation, Habit, Behavioral Intention, and Personal Innovativeness (Venkatesh, 2012). Cronbach's α for this scale is 0.95.

### 3.3.2 Pre- and Post-tests

Students completed a pre-test before the course and a post-test after completing the course. The two tests were parallel forms, each consisting of 60 multiple-choice questions designed by the course instructor and a teaching assistant.

### 3.3.3 Dialogue analysis

The dialogue data was automatically recorded by the MAIC system, with each data entry containing the following attributes: user_id, module, created_time, speaker, and content. A total of 75,896 data entries were collected, among which 4,008 were from human students.

To gain deeper insights into the learning characteristics behaviors exhibited in the dialogues and the interaction patterns between students and AI agents, the research team conducted a comprehensive coding of all dialogue data with the assistance from LLMs. The coding framework was iteratively developed and refined through multiple rounds of design, pilot testing, and expert review to ensure validity and reliability.

**Design of the coding scheme**

This study draws on the Flanders Interaction Analysis System (FIAS) and learning engagement theories to propose a preliminary coding framework comprising three dimensions for two roles. The actions of teachers and students are coded primarily based on the ten categories proposed by FIAS. After establishing the preliminary coding framework, multiple rounds of trial coding were conducted to assess its applicability. Three senior education researchers served as coders and collaboratively carried out three rounds of trial coding. In each coding trail, coders discussed their understanding of the coding scheme and made adjustments to improve the scheme. After three rounds of coding trials, the coding framework had been established as shown in Table 1. Based on this scheme, three coders conducted a trial coding based on 200 dialogue entries. The inter-coder reliability reached 0.92, and all disagreements were resolved through discussion. The differences in coding were primarily attributed to individual judgment rather than issues with the coding scheme. Therefore, the final evaluation of the coding scheme indicates that it is robust and suitable for analyzing the interactions between students and AI agents within the MAIC system.

**Table 1**

*Coding framework of the dialogue between student and AI agents*

| Role | Code |
|---|---|
| Teacher | TB1: Ask Questions; <br> TB2: Respond to questions; <br> TB3: Set plans or give directions; <br> TB4: Monitor and regulate progress; <br> TB5: Regulate emotion; <br> TB6: Others |
| Student | SB1: Ask questions; <br> SB2: Respond to questions; <br> SB3: Initiate ideas; <br> SB4: Negotiate and Confirm ideas; <br> SB5: Monitor and regulate progress <br> SB6: Manage classroom partners <br> SB7: Share emotion <br> SB8: Others |

**Coding methods**

Given the large volume of dialogue data, this study employed large language models (LLMs) to assist in the coding process. Recent research has demonstrated the feasibility and effectiveness of using LLMs to support qualitative coding in educational contexts (Long et al., 2024). To achieve high accuracy and reliability in LLM-coding, this study utilized two models (GPT-4-turbo and GLM-4) with two types of prompting strategies (few-shotting and zero-shotting). The example prompts used for LLM-based coding is provided in Appendix A.

To evaluate the accuracy of LLM-coding, a set of 200 dialogue entries that had been manually coded and reconciled by expert coders were used for LLM-coding trials. The outputs generated by the LLMs were compared with the human-coded results, yielding an inter-rater reliability of 0.81. Human coders reviewed and discussed the discrepancies, adjusted the prompts, and then re-encoded the data using the LLMs. This refinement improved the inter-rater reliability to 0.92. To further validate coding consistency, an additional 200 dialogue entries were coded by both LLMs and human coders. The inter-rater reliability remained consistently above 0.92, confirming the robustness of the LLM-based coding approach. Based on these findings, the two LLMs with few-shotting were used to code the entire dataset. Subsequently, two independent human coders systematically reviewed and reconciled all discrepancies in the LLM outputs.

As a final validation step, one human expert independently recoded the entire subset of student dialogues and compared the results with the validated LLM coding outputs. The inter-rater reliability between the human coder and the machine coding reached 0.97, indicating a high degree of agreement.

### 3.4 Data analysis

Based on the coded dialogue data, descriptive analyses were first conducted to examine the overall patterns of student-AI interaction. Sequences of behavioral codes were then mined to investigate transitions between different classroom participants (human students and AI peers) across knowledge points within each module. The frequencies of pairwise behavioral transitions were calculated to reveal interactional tendencies.

To further examine the temporal structure of interaction, lag sequential analysis was employed. This statistical method assesses the conditional probability of one behavior following another, allowing for the identification of significant behavior sequences. Through this approach, the dynamic patterns of student-AI dialogue were revealed.

In addition, group comparisons were conducted to examine differences in interaction patterns between students with high and low prior knowledge. Finally, paired-sample t-tests and one-way ANOVA were used to analyze changes in students' academic performance and attitudinal measures before and after the course.

## 4. Results

### 4.1 Student distinct behavioral engagement patterns in the multi-agent learning system

#### 4.1.1 Number and nature of student-AI dialogues

In this study, we collected and analyzed the complete dialogue history from all six modules of the course, including both student-initiated questions and AI responses. A total of 75,896 entries were recorded, of which 19,365 entries were classified as interaction data, while the remaining entries represented instructional content (i.e., AI teacher explaining course content presented on slides).

Within the interaction data, 4,005 utterances were produced by human students, accounting for 20.68% of the total. On average, each human student sent approximately 6 messages in the first two modules (6 and 6.05 respectively). The number of messages declined in later modules, with an average 3.61 in Module 4 and 3.70 in Module 6. The most active student sent 38 messages in a single module (Module 1). In general, each student message received multiple responses from multiple AI agents. The distribution of interaction counts across the six modules is shown in Figure 3. Examples of student-AI dialogue can be found in Table 2.

**Figure 3**

*Distribution of Average Interaction Counts Across Six Modules*

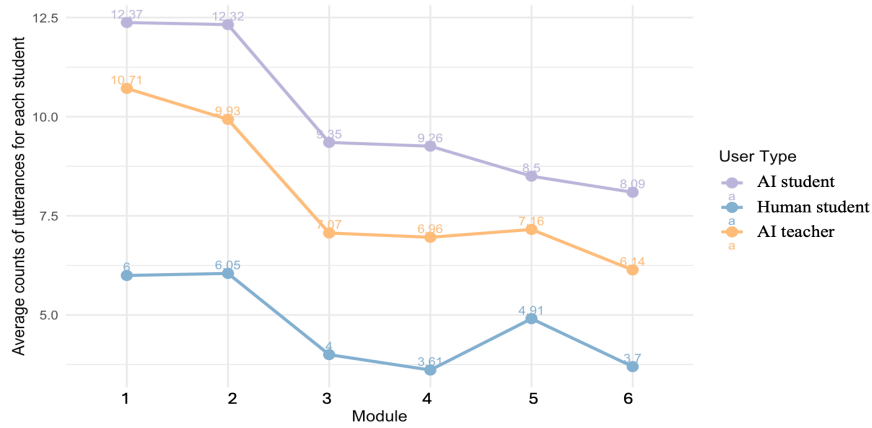

**Table 2**

*Example of conversation between student and AI agents*

| Human student | What are Residual Networks? |
|---|---|
| AI teaching assistant | Residual Networks (ResNet) are a type of deep learning model particularly suited for training very deep neural networks. By introducing the concept of "residual learning," they address the vanishing or exploding gradient problems that often occur in deeper networks. In ResNet, each residual block allows the input to bypass several layers and be directly added to the block's output. This promotes the flow of information in deep networks, enhancing both training efficiency and model performance. ResNet achieved groundbreaking results in various visual recognition tasks and won the championship in the 2016 ImageNet competition. |
| AI teacher | Residual networks are a major innovation in deep learning. By introducing "skip connections," they address the vanishing gradient problem in deep networks, allowing the network to go deeper and learn more effectively. The design of the residual block enables each layer to learn the difference—i.e., the "residual"—between its input and output. This concept effectively doubles the number of layers in the network and boosts model performance. Keep up the good work, and you'll make even more discoveries! |
| AI peer-Questioner | That's amazing! Thank you for explaining the concept of Residual Networks, Teacher! I am wondering, can the architecture of ResNet be applied to other types of deep learning tasks? What do other classmates think? |

The distribution of engagement behaviors by different roles is presented in Table 3. As shown in the table, nearly half of the utterances by human students (48.19%) were asking questions, followed by monitoring and regulating progress (14.09%). There was also a notable proportion of initiating ideas (13.84%). Utterances related to negotiating and confirming ideas, managing classroom partners, and sharing emotions were less frequent, accounting for 6.87%, 5.66%, and 4.91%, respectively. For the AI Teacher, over half of their behaviors during interactions were responding to students (52.51%). The behaviors of AI peers were more concentrated on "initiate ideas", accounting for about 36.86%, followed by asking questions (20.59%) and negotiating ideas (19.03%).

**Table 3**

*The distribution of engagement behaviors by different roles*

| Role | Behavior | Count | Percentage |
| --- | --- | --- | --- |
| **Human student** | Initiate ideas | 665 | 13.84% |
| | Respond to questions | 156 | 3.24% |
| | Ask questions | 2316 | 48.19% |
| | Negotiate and Confirm ideas | 330 | 6.87% |
| | Monitor and regulate progress | 677 | 14.09% |
| | Manage classroom partners | 272 | 5.66% |
| | Share emotion | 236 | 4.91% |
| | others | 154 | 3.20% |
| **AI classmate** | Initiates ideas | 5200 | 36.86% |
| | Respond to questions | 1996 | 14.15% |
| | Ask questions | 2904 | 20.59% |
| | Negotiate and Confirm ideas | 2684 | 19.03% |
| | Monitors and regulates progress | 33 | 0.23% |
| | Manage classroom partners | 13 | 0.09% |
| | Share emotion | 1243 | 8.81% |
| | others | 33 | 0.23% |
| **AI teacher** | Respond to questions; | 6824 | 52.51% |
| | Ask Questions | 284 | 2.18% |
| | Set plans or give directions; | 1572 | 12.10% |
| | Monitors and regulates progress; | 52 | 0.40% |
| | Regulate emotion; | 4223 | 32.50% |
| | Others | 40 | 0.32% |

To better delineate students' unique engagement patterns within the multi-agent learning environment, we compare students' engagement between the multi-agent setting and a MOOC context equipped with a Chatbot for Q&A support. On average, students in the MOOC setting engaged with the chatbot 1.38 times during each module, significantly fewer than the interactions observed in the multi-agent environment.

**4.1.2 Behavioral Transition Patterns in Student-AI Interactions**

To uncover patterns of student engagement in interactions with multiple AI agents, Transition Analysis and LSA were employed. The heatmap presented in Figure 4 visualizes the behavioral transitions among different classroom participants, including the AI teacher, AI peers and human students. Each cell represents a transition between two behavior types, with darker colors indicating higher transition frequencies.

**Figure 4**

*Heatmap of behavioral transitions between different roles*

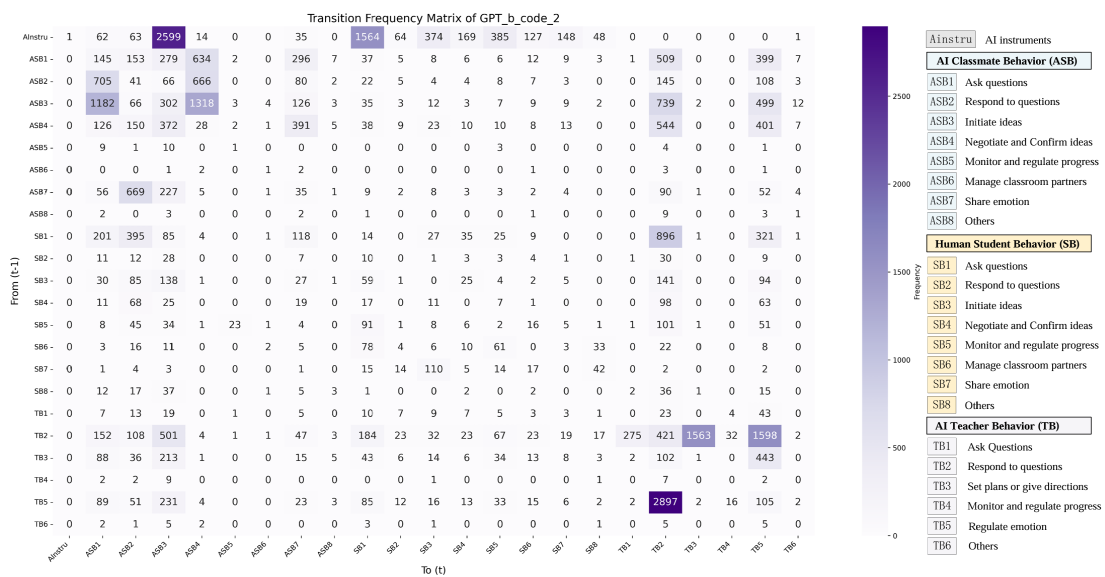

As shown in Figure 4, frequent internal behavioral transitions within specific AI roles could be observed, such as ASB3 -> ASB1 (AI students firstly initiate ideas, then they ask others' opinions), ASB3 -> ASB4 (AI students initiate their own ideas, then negotiate and confirm ideas with others ); and ASB2 -> ASB1 (AI students firstly response to others, then ask follow-up questions), ASB2 -> ASB4 (AI students response to others, then negotiate ideas with others ). Similarly, internal behavioral transitions within the AI teacher also occur frequently, such as TB5->TB2 (AI teacher first regulates others' emotions and then asks questions); TB2-> TB3 (AI teacher first asks questions then gives some instructions).

Regarding the behavioral transition between different roles, some frequent transitions patterns were observed:

**(1) Instruction -> SB1, SB3, SB5**: After instruction, students are likely to ask questions (SB1) (28%), Initiate ideas (SB3) (7%), or monitor and regulate progress (SB5), such as by prompting the teacher to continue or revisit the material (7%).

**(2) TB2 -> SB1:** Following a teacher's response to a student (TB2), students often ask follow-up questions (SB1) (4%).

**(3) SB1 -> TB2, TB5, ASB2, ASB1:** When a human student asks a question (SB1), more than half the time, it is the AI Teachers that give responses to students (42%) and praise students (15%). Additionally, AI students may also respond (19%) or pose further questions based on what human students asked (4%).

**(4) SB3 -> TB2, ASB3:** After a human student initiates ideas and spontaneously shares their opinion (SB3), the AI Teacher often responds to the human students, acknowledging the student's ideas and providing feedback (23%). Similarly, AI peers may seize this opportunity to share their perspectives, contributing to a more diverse classroom dialogue (23%).

**(5) SB5 -> TB2**: When a human student manages the classroom progress, such as asking the teacher to continue to the next topic or go back to the previous topic, usually the AI teacher gives a direct response (25%) such as introducing the new topic or explaining more about the previously taught topics.

To further capture the behavioral sequences that characterize richer student-AI interactions, LSA was conducted. The results, presented in Figure 5, highlight meaningful interaction sequences both within roles and between roles. In particular, six significant behavioral chains involving all three types of classroom participants—the AI teacher, AI peers, and human students—were identified. These sequences represent statistically significant transition patterns and illustrate how engagement is distributed across multi-agent interactions.

**(1) TB1 -> SB2 -> SB6 -> ASB6 ->ASB6/SB6:** The AI teacher asked a question, followed by a human student providing an answer, after which the student tended to invite specific AI agents to respond, and subsequently, an AI student also extended an invitation to other AI agents or the human student for further engagement.

**(2) TB1 -> SB2 -> SB6 ->SB4 -> ASB2：** The AI teacher asked a question, a human student responded, and then the student invited other AI agents to reply. The human student further negotiated or sought confirmation about their ideas, after which the AI student responded.

**(3) TB1 -> SB2 -> SB6 -> SB1 -> TB2/ASB2：** The AI teacher asked a question, a human student responded, and then the student invited the AI to reply. After receiving the response, the human student further asked a question, to which either the teacher or the AI student provided a response.

**(4) TB1 -> SB3 -> SB4 -> ASB2：** The AI teacher asked a question, a human student shared their perspective, and then the student sought confirmation, to which the AI student engaged in the discussion.

**(5) TB1 -> SB4 -> ASB2：** The AI teacher asked a question, a human student negotiated or sought consultation, and the AI student participated in the discussion.

**(6) SB7 ->SB5 -> SB1 -> TB2 -> TB4 -> ASB2**：The student shared their feelings, then regulated the process (e.g., by advancing the course progress), and initiated questions. The teacher responded and further regulated the process, after which the AI student provided a reply.

Based on the significant behavior transition sequences listed above, we have identified two main patterns: co-construction of knowledge and co-regulation. These two patterns may sometimes stand alone, as in (TB1 -> SB4 -> ASB2), but often they can appear in combination, such as starting with co-construction of knowledge (TB1 -> SB2) followed by co-regulation (SB6 -> ASB6 ->ASB6/SB6). In addition to the interaction patterns involving all three roles presented above, we also observed that the direct interaction between human students and AI peers primarily consisted of the AI students responding to questions initiated by the human students (SB1->ASB2). However, the AI students' responses did not significantly trigger follow-up discussion or other activities from the human students. Interestingly, when it came to regulating learning progress and coordinating with class partners, the interactions between AI peers and human students were both bidirectional and statistically significant, suggesting a balanced exchange of regulatory behaviors.

This interaction pattern between student-AI peers stands in contrast to the student-AI teacher interaction, where the dominant form of engagement was cognitive in nature, such as question-and-answer exchanges. The differentiation highlights how AI agents with different roles elicit distinct types of engagement from students in a multi-agent learning environment.

**Figure 5**

*LSA result of behavioral interaction between roles*

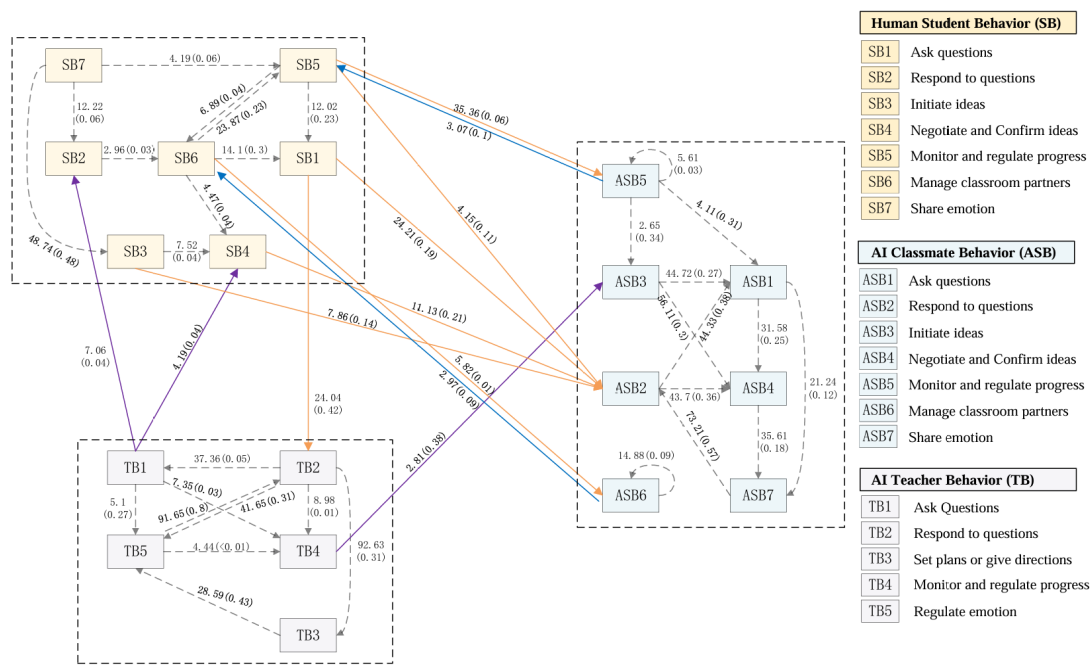

**4.2 Comparing engagement patterns between students with different prior attainment levels**

To examine how prior knowledge influences students' engagement and interaction patterns, we categorized participants based on their pre-test performance. Specifically, the top 30% of students were classified as the high prior knowledge group, while the bottom 30% were categorized as the low prior knowledge group. This percentile-based approach ensured a sufficient number of students in each group for meaningful comparison. As a result, 33 students were included in the high prior knowledge group ($M = 8.52$, $SD = 0.66$), and 38 students were assigned to the low prior knowledge group ($M = 4.49$, $SD = 0.75$).

The average number of messages sent by the high prior knowledge group ($M = 35.76$, $SD = 46.74$) was significantly higher than the low prior knowledge group ($M = 14.47$, $SD = 23.84$), $z = -2.750$, $p = .006$. In addition, students in the high-prior-knowledge group engaged in significantly more multi-turn conversations with AI agents than their low-performing peers, $z = -2.83$, $p = .005$. Since students with higher knowledge level sent a greater number of messages, the corresponding responses from AI agents were also more frequent, resulting in a significantly denser interaction dynamic. Specifically, the average number of messages generated by the AI peers in the high prior knowledge group was 80.06 ($SD = 85.35$), which was significantly higher than that in the low prior knowledge group ($M = 30$, $SD = 28.25$), $z = -3.033$, $p = .002$. Similarly, the AI teacher responded more frequently to students in the high prior knowledge group ($M = 54.82$, $SD = 48.91$) than in the low prior knowledge group ($M = 27.32$, $SD = 26.56$), $z = -2.866$, $p = .004$.

Regarding students' behaviors, compared to those with lower prior knowledge, students with higher prior knowledge exhibited a higher proportion of behaviors involving the active initiation of their own ideas (15.5% vs. 12.1%). Additionally, students with higher prior knowledge demonstrated significantly more regulatory behaviors, including monitoring and regulating the learning process (12.9% vs. 11.2%), managing AI peers (7.5% vs. 4.5%), and sharing emotions (7.4% vs. 2.6%). In contrast, students with lower prior knowledge were more likely to engage in question-raising behaviors (55.3% vs. 45.2%) and negotiating and confirming ideas (8.9% vs. 5.7%).

**Figure 6**

*The proportional distribution of behaviors of the two groups of students*

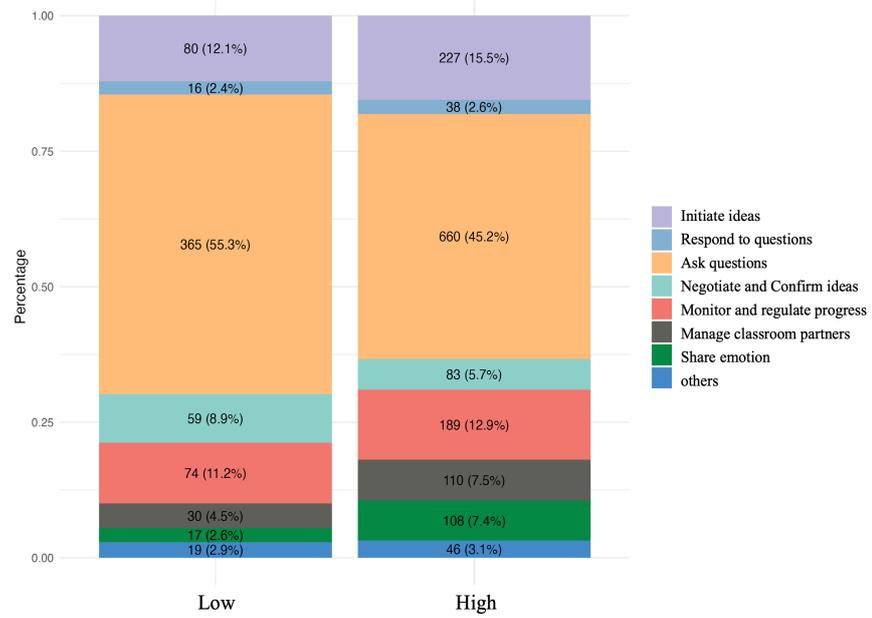

To further explore the differences in behavioral engagement patterns between the two groups, we used LSA to analyze the behavioral transition sequences in each group. As illustrated in Figure 7, the group with higher prior knowledge exhibited a greater number of significant behavioral sequences compared to the low prior knowledge group. Specifically, in the interactions between high prior knowledge students and AI agents, the monitoring and regulating process (SB5) plays a more central role, being involved in many significant sequences. This includes some internal behavioral transitions, such as **SB7 –> SB5**, and **SB6 –> SB5**. This behavior also plays a part in external interactions, such as regulating the course pace based on the AI teacher's directions or AI peers' monitoring and regulating, which in turn triggers the regulatory behaviors of AI peers (**TB3/ASB5 –> SB5 –>ASB5/ASB6**).

In comparison, the group with lower prior knowledge has more significant sequences involving negotiating and confirming behaviors (SB4). Significant sequences include initiating ideas then negotiating or confirming ideas (SB3 –> SB4), managing classmates and negotiating ideas (SB6 –> SB4), or sharing emotions then confirming ideas (SB7 –> SB4). These sequences subsequently triggering responses from AI peers (**SB3/SB6/SB7 –> SB4 –>TB2/ASB2**). In contrast, the negotiating and confirming behaviors (SB4) in the group with higher prior knowledge mostly follow the behavior of responding to questions, leading to further responses from the AI teacher and AI peers (**SB2/SB6 –> SB4 –>TB5/ASB2**).

**Figure 7**

*Figure 7a: LSA result of behavioral interaction between students with higher prior knowledge level and AI agents*

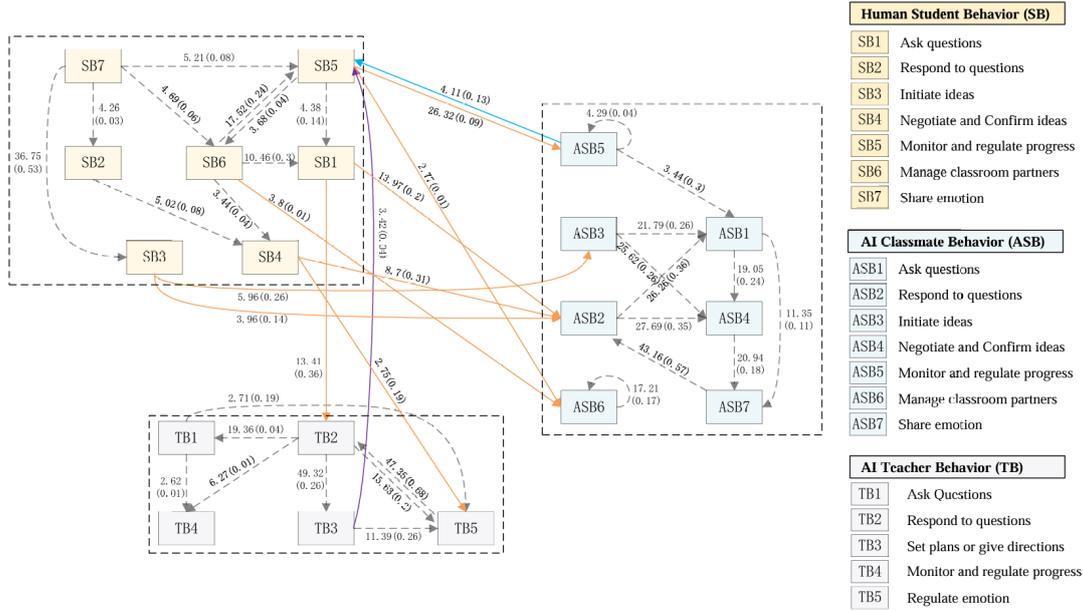

*Figure 7b: LSA result of behavioral interaction between students with lower prior knowledge level and AI agents*

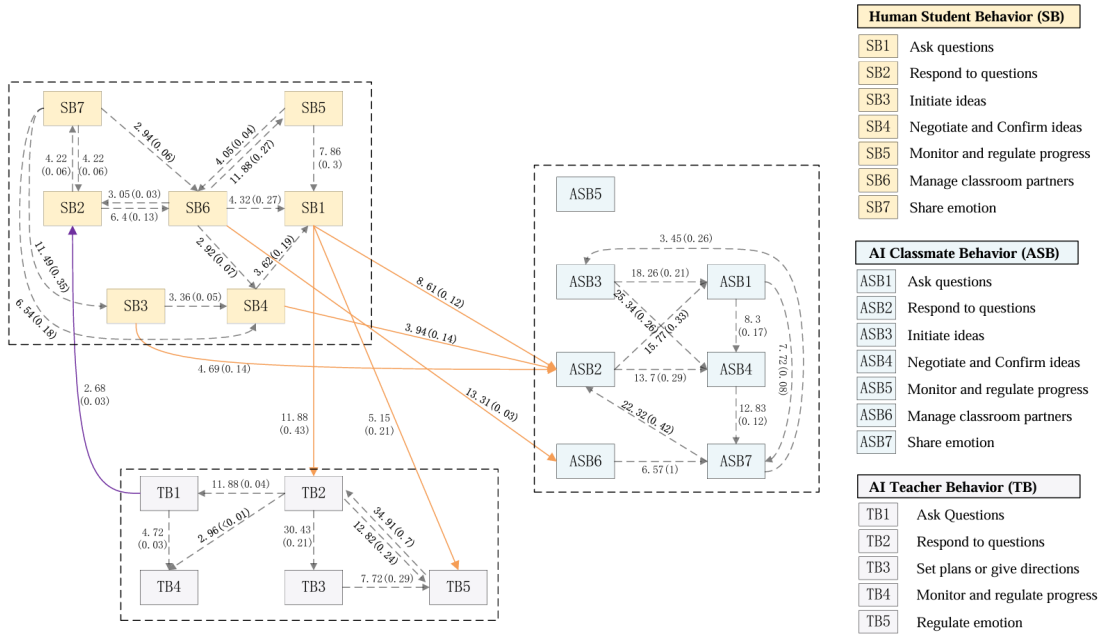

### 4.3 Learning Gains and Motivational Shifts in Multi-Agent Environments

To examine students' learning gains throughout the course, a paired-sample t-test was conducted to compare pre-test and post-test performance. The results revealed a statistically significant improvement in students' scores from the pre-test ($M$ = 6.45, $SD$ = 1.61) to the post-test ($M$ = 6.77, $SD$ = 1.38), $t(109)$ = –2.54, $p$ = .013. In addition to academic performance, students' technology acceptance of AI also showed a significant increase. The average score on the AI acceptance scale rose from $M$ = 4.60 ($SD$ = 0.60) before the course to $M$ = 4.73 ($SD$ = 0.52) after the course, $t(105)$ = 3.13, $p$

= .001. A closer examination of the subdimensions of AI technology acceptance revealed that the observed increase was particularly significant in two key areas: Effort Expectancy and Facilitating Conditions. For the Effort Expectancy, students reported a significantly more favorable perception after completing the course ($M$pre = 5.71, $SD$ = 1.10; $M$post = 6.03, $SD$ = 0.80), $t(105)$ = 3.66, $p < .001$). Similarly, the Facilitating Conditions dimension also showed a significant improvement ($M$pre = 5.57, $SD$ = 0.96; $M$post = 5.86, $SD$ = 0.88), $t(105)$ = 3.13, $p$ = .001).

To further examine how prior knowledge influenced students' learning outcomes, a one-way ANOVA was conducted to compare learning gains across students with different levels of prior knowledge. The results revealed a significant main effect of prior knowledge level on learning gain, $F(2, 69)$ = 22.95, $p < .001$, partial $\eta^2$ = .400. Students in the low prior knowledge group achieved the greatest learning gains ($M$ = 1.88, $SD$ = 1.36), followed by the medium group ($M$ = 0.36, $SD$ = 1.34), while the high group experienced a learning loss ($M$ = –1.09, $SD$ = 1.37). A visual comparison of the pre- and post-test results among the groups is provided in Figure 8

**Figure 8**

*Differences in Learning Gains Across Prior Knowledge Groups*

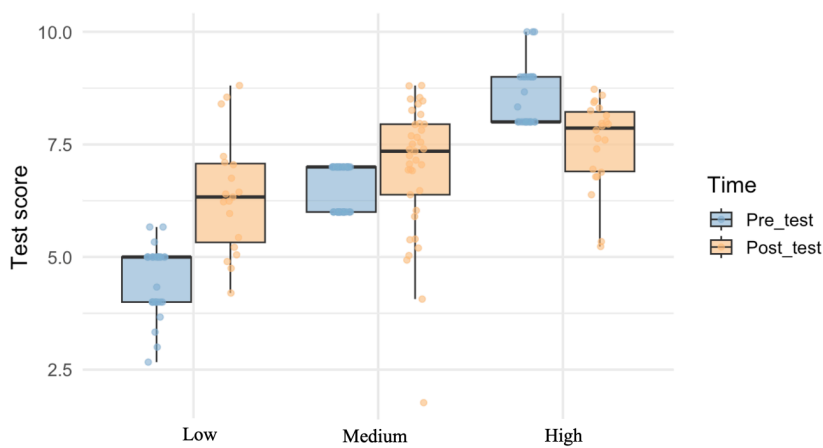

In addition to cognitive learning outcomes, we also examined changes in students' non-cognitive factors, including learning motivation and technology acceptance. A one-way ANOVA revealed no significant differences in pre-course AI technology acceptance, $F(2, 118)$ = 0.815, $p$ = .445. This suggests that students' initial attitudes toward AI technology were statistically equivalent across groups. Similarly, no significant group differences were found in pre-course learning motivation across the three prior knowledge groups, $F(2, 118)$ = 0.103, $p$ = .902, indicating that students began the course with comparable motivational levels regardless of their prior knowledge level. Using pre-course AI technology acceptance as a covariate, the ANCOVA showed that after adjusting for the covariate, there was no significant difference in post-course technology acceptance among the three prior knowledge groups, $F(2, 95)$ = 2.30, $p$ = .106. To investigate whether students' post-course learning motivation varied across different prior knowledge groups, we conducted an ANCOVA using pre-course learning motivation as a covariate. The results revealed a significant effect of prior knowledge group on post-course learning motivation, $F(2, 95)$ = 4.15, $p$ = .019, after adjusting for the covariate.

Pairwise comparisons of the adjusted means showed that students in the low prior knowledge group reported significantly higher post-course motivation ($M = 1.56$, $SE = .10$) than those in the high prior knowledge group ($M = 1.14$, $SE = .10$), with the medium group falling in between ($M = 1.36$, $SE = .08$).

**5. Discussion**

**5.1 Learning Mechanisms in Multi-Agent Human-AI Interaction**

This study investigated how students interacted with multiple AI agents, each assigned distinct pedagogical roles—such as AI teachers and AI peers—within an LLM-empowered interactive learning environment. Through fine-grained analysis of behavioral sequences, two main patterns of engagement were identified: knowledge acquisition and co-construction, and co-regulation of the learning process. Each pattern reflects a fundamental mechanism by which students collaboratively constructed meaning with AI agents and managed their learning paths.

The knowledge acquisition and co-construction pattern is typically initiated by students when they express queries about specific knowledge points. This was subsequently addressed by AI agents through direct responses, elaborations, or prompting further inquiries. These interaction sequences reflect behaviors aligned with cognitive engagement (Fredricks et al., 2004) and resemble traditional forms of instructional scaffolding, enabling students to progressively deepen their understanding. In contrast, the co-regulation pattern emerged when students proactively invited AI agents into the dialogue, monitored progress, or negotiated direction. These behaviors align with the conceptualization of socially shared regulation of learning (SSRL) (Hadwin et al., 2011), where regulation is co-constructed in real-time through shared interaction.

Importantly, these patterns were not evenly distributed across learners. Students with higher prior knowledge exhibited more frequent and agentive co-regulation behaviors, suggesting stronger metacognitive control and a greater capacity to orchestrate the learning process (Negi et al., 2022; Stanton et al., 2021). Meanwhile, students with lower prior knowledge tended to rely more on cognitive support sequences, engaging with the AI agents primarily to receive guidance or explanations. This differentiation highlights how learners position themselves differently in relation to AI agents based on their own levels of confidence, control, and perceived need.

These findings provide empirical grounding for extending existing frameworks in computer-supported collaborative learning, especially those emphasizing the interdependence of collaboration, regulation, and knowledge co-construction as central to productive learning (Stahl, 2006; Hadwin et al., 2011). While such mechanisms have traditionally been studied in human-human interaction settings, our results suggest that similar structures can emerge in human-AI dialogues—albeit with new interactional forms.

**5.2 Knowledge gains from interactive learning**

Students with low prior knowledge achieved significantly higher learning gains after completing the course, suggesting that less-prepared learners benefited more. According to Cognitive Load Theory (Sweller, 1988), novices typically experience high extraneous cognitive load when grappling with new material in the absence of established schemas. students with low prior knowledge demonstrated more frequent knowledge-seeking and confirming behaviors, which were met with responses from multiple AI agents. This interaction pattern helped reduce extraneous processing by providing on-demand guidance and feedback, thereby freeing the students' working memory capacity for germane load – the constructive mental effort used to form new schemas. As a result, less-prepared learners could focus on integrating new information into their existing understanding, facilitating effective schema construction and leading to greater performance improvements. Additionally, from the perspective of Distributed Cognition (Pea, 1993), the AI agents functioned as part of an expanded cognitive system for the students, allowing learners to offload some of the cognitive burden onto the technology. Such distributed intelligence is especially advantageous for novices, as it augments their effective cognitive capacity and supports the development of complex knowledge structures that they might struggle to build independently.

In contrast, students with higher prior knowledge showed relatively limited learning gains, a potential explanation being ceiling effects (Douglas et al., 2020; Yuan et al., 2021). A ceiling effect occurs when learners begin an intervention already near mastery, leaving little room for measurable improvement (Staus et al., 2021). As these students entered the course with strong initial scores, so even if they learned something, the assessments may not have captured those small increments. From a cognitive load standpoint, the instructional support that greatly aided novices provided diminishing returns for experts. In fact, overly detailed guidance can become redundant or even counterproductive for advanced learners – an instance of the expertise reversal effect. Consistent with this, participants with higher prior knowledge were observed to use the AI tools not only for cognitive activities but also in a regulatory manner, primarily for course navigation and regulation. For example, these students often skipped familiar sections, fast-forwarded through practice exercises, or used the AI to quickly confirm what they already knew. These behaviors suggests that the pedagogical support provided by AI agents in the system was underutilized by these learners. Furthermore, it is possible that the course itself added little beyond what these students could accomplish on their own, thus yielding minimal visible gains.

**5.3 Non-cognitive gains from interactive learning**

In addition to students' cognitive gains, we also examined changes in non-cognitive aspects. Our results showed that students' acceptance of AI technology improved across the board, regardless of prior knowledge. According to TAM, when a technology is both easy to use and demonstrably helpful, users develop a more favorable attitude and stronger intention to use it in the future (Davis, 1989; Scherer et al., 2019). This is exactly what we observed: the AI-empowered multi-agent learning environment improved students' effort expectancy (they could accomplish tasks with less work) and provided clear facilitation of the technology usage, which together fostered greater acceptance. However, an important question remains as to what caused the increase in technology acceptance – was it the experience of learning with multiple AI agents, or the nature of the course content being about artificial general intelligence? In our study, these two factors were intertwined. All participants were enrolled in the course featuring AI-based instruction and covering AI-related topics, so their growing comfort with the technology could be due to either (or both) of these elements. Future research might isolate these factors by, for example, comparing technology acceptance in courses that use an AI-supported system for non-AI topics versus courses about AI that do not use AI tools in teaching.

Moreover, we analyzed students' learning motivation and found that students with low prior knowledge reported significantly higher motivation after the course compared to their peers with higher prior knowledge. We interpret this result through the lens of Expectancy-Value Theory (Eccles & Wigfield, 2002), which posits that student engagement is shaped by their expectations of success and the perceived value of the learning task. Although students with higher prior knowledge likely entered with high expectations of success due to their existing competencies, this very confidence might have diminished their perceived value of fully engaging with the course content. Specifically, if these students anticipated a challenging and intellectually stimulating experience but found the material easier than expected, the discrepancy could have negatively impacted their motivation. Conversely, students with low prior knowledge benefited substantially from the cognitive scaffolding provided during the learning process, experiencing notable improvements in their understanding. This tangible sense of accomplishment likely heightened their motivation, further encouraging their active participation and engagement.

**5.4 Practical Implications**

This study offers several implications for the design of multi-agent AI learning systems. The results suggest that differentiated support, real-time orchestration, and stronger peer-agent design are essential for realizing the full potential of multi-agent supported learning environments.

First, agent roles should be adaptive to learners' prior knowledge. For low prior knowledge students, AI agents should provide scaffolded support—offering step-by-step guidance and content clarification. In contrast, high prior knowledge students benefit more from reflective prompts that elicit reasoning, deep elaboration, or perspective-taking. Static, one-size-fits-all agent behaviors risk under-supporting novices and disengaging advanced learners.

Second, agent participation should be dynamically regulated based on students' interaction patterns. The system can monitor cues such as frequent help-seeking or surface-level responses to adjust agent involvement in real-time. For instance, when students demonstrate confidence and fluency, teacher agent interventions can be minimized, allowing peer agents or self-directed exploration to take precedence.

Third, the design of peer-like agents requires further refinement. In this study, AI peers rarely triggered sustained follow-up or elaboration from students. This suggests a need for improved social presence, epistemic responsiveness, and dialogic scaffolding. Enhancing peer agents' ability to engage in meaningful co-construction could increase their effectiveness in simulating collaborative learning.

**5.5 Limitations**

Several limitations should be acknowledged in this study. First, the sample was drawn from a single course at one university, which limits the generalizability of the findings. Future research should examine more diverse populations and learning contexts. Second, the analysis focused on the structural coding of dialogues, capturing behavioral patterns but not the semantic or discursive content. Integrating semantic or sentiment analysis could offer deeper insights into the quality and meaning of

student–AI interactions. Third, the post-test was administered immediately after course completion, capturing short-term learning gains only. Future studies should incorporate delayed assessments or transfer tasks to examine long-term retention and knowledge application.

Despite these limitations, the study provides a valuable starting point for understanding engagement dynamics in multi-agent learning environments and highlights directions for more comprehensive future inquiry.

**6. Conclusion**

This study investigated how students interact with multiple AI agents in an LLM-empowered interactive learning environment and how such interactions shape their cognitive and motivational outcomes. By analyzing over 19,365 lines of dialogue data and comparing behavioral engagement, learning gains, and motivational changes across students with different levels of prior knowledge, the study offers new insights into the mechanisms and effects of human–AI collaboration in educational contexts. Two core engagement patterns—co-construction of knowledge and co-regulation—were identified as central to the learning process, with students dynamically navigating these patterns based on their needs and prior knowledge. Notably, students with lower prior knowledge exhibited significantly greater learning gains, suggesting that AI-supported interaction may help narrow learning gaps by offering personalized scaffolding. In contrast, students with higher prior knowledge demonstrated more autonomous regulation behaviors but benefited less in terms of measurable outcomes, possibly due to ceiling effects and reduced challenge.

The findings extend existing computer-supported collaborative learning and self-regulated learning theories into AI-mediated environments, highlighting the potential of multi-agent systems to support both individualized learning paths and distributed engagement mechanisms. At the same time, the study raises important considerations for the design of future AI-supported learning systems, including adaptive agent roles, real-time interaction orchestration, and more effective peer-like AI design. This study offers a foundational step toward understanding how learners engage with—and benefit from—multi-agent AI in real-world educational settings. Future work should continue to explore the semantic depth of AI-student interactions, assess long-term learning outcomes, and refine agent design to optimize both cognitive and motivational support for diverse learners.